\renewcommand{\thefootnote}{\fnsymbol{footnote}}

\newcommand{\be}{\begin{equation}}
\newcommand{\ee}{\end{equation}}
\newcommand{\bea}{\begin{eqnarray}}
\newcommand{\eea}{\end{eqnarray}}
\newcommand{\bfig}{\begin{figure}}
\newcommand{\efig}{\end{figure}}

\newcommand{\bran}{KBBD}
\newcommand{\branc}{KBBD}
\newcommand{\tbar}{\mbox{$\langle t \rangle$}}
\newcommand{\bdelt}{\mbox{$\Delta t$}}
\newcommand{\tm}{\mbox{$t_{\rm max}$}}
\newcommand{\delt}{\mbox{$\delta t$}}
\newcommand{\ld}{l_d}
\newcommand{\cis}{{\em cis\/}}
\newcommand{\trans}{{\em trans\/}}
\newcommand{\alysin}{$\alpha$-hemolysin}

\newcommand{\cf}{\mbox{${\cal F}$}}
\newcommand{\lfp}{\mbox{${\cal L}$}}
\newcommand{\unr}{\mbox{$u_n^{\rm R}(k,x)$}}

\newcommand{\unlo}{\mbox{$u_n^{\rm L}(k,x_0)$}}
\newcommand{\lamtext}{\mbox{$\lambda_n(k)$}}
\newcommand{\lam}{\lambda_n(k)}

\newcommand{\ur}{u^{\rm R}}
\newcommand{\ul}{u^{\rm L}}
\newcommand{\statold}{\mbox{$u_0^{\rm R}(k=0,x)$}}
\newcommand{\psinr}{\mbox{$ \psi_n^{\rm R}(k,x)$}}
\newcommand{\ione}{\mbox{$I_1^{(0)}$}}
\newcommand{\itwo}{\mbox{$I_2^{(0)}$}}
\newcommand{\mydel}{\mbox{$(1-e^{-Fa/\kt})$}}
\newcommand{\uoo}{\mbox{$u_0^{{\rm R},1}(x)$}}
\newcommand{\kb}{\mbox{$k_{\rm B}$}}
\newcommand{\kt}{k_{\rm B} T}
\newcommand{\tz}{\mbox{$t_{\rm Z}$}}
\newcommand{\tp}{\mbox{$t_{\rm pore}$}}
\newcommand{\up}{\mbox{$u_{\rm p}$}}
\newcommand{\ui}{\mbox{$u_{\rm i}$}}
\newcommand{\lch}{\mbox{$l_{\rm ch}$}}

\def\gta{\lower.3ex\hbox{\ \rlap{$^{\displaystyle>}$}$_{\displaystyle\sim}$\ }}
\def\lta{\lower.3ex\hbox{\ \rlap{$^{\displaystyle<}$}$_{\displaystyle\sim}$\ }}

\documentstyle[aps,twocolumn,harvard]{revtex}

\citationstyle{dcu}  

\begin{document}

\title{Driven Polymer Translocation Through a Narrow Pore}
\author{David K. Lubensky and David R. Nelson}
\address{Department of Physics, Harvard University, Cambridge MA  02138}
\date{\today}

\maketitle

\renewcommand{\thefootnote}{\fnsymbol{footnote}}

\begin{abstract}
Motivated by experiments in which a polynucleotide is
driven through a proteinaceous pore by an electric field, we study the
diffusive motion of a polymer threaded through a narrow channel with
which it may have strong interactions.  We show that there is a range
of polymer lengths in which the system is approximately
translationally invariant, and we develop a coarse-grained description
of this regime.  From this description, general features of the
distribution of times for the polymer to pass through the pore may be
deduced.  We also introduce a more microscopic model.  This model
provides a physically reasonable scenario in which, as in experiments,
the polymer's speed depends sensitively on its chemical composition,
and even on its orientation in the channel.  Finally, we point out
that the experimental distribution of times for the polymer to pass
through the pore is much broader than expected from simple estimates,
and speculate on why this might be.
\end{abstract}

\section{Introduction}
\par
Modern polymer physics has achieved great success with models in which
the polymer is regarded as a flexible, uniform ``string'' whose
conformational entropy dominates the system's
behavior~\cite{deg-bk,doi-ed}.  Although this is usually an excellent
description, in some situations other interactions can become
important.  One example is the insertion of a polymer
into a pore of diameter comparable to the size of the chemical repeat
units that make up the polymer.  Although perhaps unusual with
synthetic polymers, such a situation can easily occur in biological
systems.  For example, Kasianowicz, Brandin, Branton, and Deamer
(hereafter KBBD) have recently detected single strands of RNA
(polyuridylic acid) passing through a 1.5 nm pore formed by a
membrane-bound protein~\cite{branton}.  Szab\`{o} and
coworkers~\cite{szabo,szabob} and Hanss and coworkers~\cite{hanss}
have studied similar systems.  In addition to their intrinsic
interest, these experiments may eventually lead to a single-molecule
RNA and DNA sequencing technique.  
More generally, most cells must transport macromolecules across
membranes in order to function; in several cases, relatively
``thick'' molecules are believed to pass through
nanometer-scale channels.  The translocation of polynucleotides
through proteic pores has been implicated in a variety of
processes, including phage infection and bacterial
conjugation~\cite{dreis}, the uptake of oligonucleotides by certain
organs~\cite{hanss}, and transport across the nuclear envelope in
plants~\cite{citovsky}.  It has been speculated that some of these transport
pathways could eventually prove important in gene therapy~\cite{szabob,hanss}.
Similarly, polypeptide-conducting channels play an important role in
protein kinesis~\cite{schatz,sblobel}; in a few instances, the
translocation may even be driven by electrophoretic
effects~\cite{aschatz}.

\par
There exists a considerable literature on the confinement of polymers
in channels of diameter significantly larger than the polymers'
persistence length~\cite{deg-bk}; well-developed scaling techniques
can be used in the theoretical treatment of this regime.  Recently,
theorists have also shown an interest in the opposite limit of a very
narrow, almost point-like hole.  For example, several groups have
studied the diffusion of polymers across idealized, infinitely thin
membranes~\cite{carl,dimarzio,yoon-deutsch,lee-obu,park-sunga,park-sungb,sung-park}.
The pore and the membrane are viewed as hard walls whose only
interaction with the polymer is steric, and the emphasis is on how the
walls' presence decreases the entropy and slows the dynamics of those
parts of the polymer outside of the hole.  Possible mechanisms for the
active transport of polymers through pores in biological systems have
also been studied~\cite{poo,spo,sung-park}.

\par
Inspired largely by the experiments of
\bran, in this paper we consider an different scenario:  We study
the motion of a homopolymer threaded through a narrow pore with which
it has strong interactions.  The pore is taken to be sufficiently
small that no more than one polymer diameter can fit in it at a given
time; in particular, ``hairpin'' bends are not allowed to pass through
the channel.  We also put aside the question of how the polymer first
enters the hole, focusing instead on the dynamics once one end has
been inserted.  We then argue that, in the presence of a force driving
the polymer through the pore, there should be a regime in which the
polymeric degrees of freedom outside of the pore can be neglected, and
the system is effectively one-dimensional.  In this limiting case we
propose a two-tiered picture: a coarse-grained ``macroscopic''
description of wide validity and a simple ``microscopic'' model from
which the ``macroscopic'' parameters may be calculated.  Our approach
follows several authors~\cite{poo,spo,park-sunga,park-sungb,sung-park}
in viewing the translocation process as essentially diffusion in one
dimension; we differ, however, in emphasizing the role that
interactions with the pore itself play in this diffusion process.  On
the more microscopic level, we include the effects of these
interactions through a tilted washboard potential, similar to models
of laser mode locking~\cite{haken} or phase dynamics in Josephson
junctions~\cite{halp-ambegok} (see figures~\ref{fig4} and~\ref{fig5}).
The periodic modulation of the potential reflects the periodicity of
the polynucleotide's sugar-phosphate backbone.  The importance of
polymer-pore interaction has previously be emphasized by Bezrukov and
coworkers~\cite{bezr}; our model also bears some similarity to work on
gel electrophoresis that examines the importance of local ``solid
friction'' forces between the polyelectrolyte and the
gel~\cite{deutsch-solid,burl-deu-a,viovy-duke,burl-deu-b,deutsch-yoon}.
Although the macroscopic parameter values for KBBD's system differ in
some respects from those predicted by our microscopic model, we are
nonetheless able to make several fairly robust predictions.  More
importantly, we show how a simple physical mechanism can account for
several striking features of the data of
\branc.  We thus hope that our work will provide a useful contribution
to our understanding of the translocation of polyelectrolytes.

\par
Since our analysis relies heavily on KBBD's results,
the next section sketches some salient features of their data.
We then introduce a long length scale
``hydrodynamic'' description of one-dimensional diffusion and use it
to calculate the distribution of passage times for a polymer being
driven through a pore.  The arguments used to arrive at these results
are quite general; in particular, they require few assumptions about
the details of the microscopic dynamics of the system.  There are,
however, circumstances when our approximations break down, and we
consider these next.  This section also serves
to emphasize several aspects of the experiments that will guide our
choice of the microscopic model in the succeeding section.  After
introducing this microscopic model, we use it to calculate a mean drift
velocity and an effective diffusion coefficient and
compare them to values estimated from KBBD's data.  These comparisons will reveal certain
features that cannot be accounted for by our model in its simplest
form, so we then discuss possible reasons for the
discrepancy, as well as touching briefly on several applications of
our calculations.  We conclude by summarizing our results and
highlighting some issues that remain open.

\section{Experimental Background}
\label{exptal}
\par
In the experiments of interest to us, \bran\ worked with a {\em
Staphylococcus aureus} $\alpha$-hemolysin ion channel in an artificial
lipid bilayer membrane (diphytanoyl-PC).  This channel has the advantage that for many purposes it may be considered
always to be open.  The
\alysin\ protein has recently been crystallized and an x-ray structure
obtained~\cite{lysin-struct}.  This reveals a mushroom-shaped complex
with a roughly 10 nm long solvent-filled channel.
The channel is 1.5 nm in diameter at its narrowest constriction,
barely larger than the diameter of a single polynucleotide strand.
  After inserting a single pore into a bilayer membrane and
applying a transmembrane potential of between 110 and 140 mV, \branc\
added homopolymeric {\em single-stranded} DNA or RNA to one side of the
membrane, designated
\cis.  The samples of polynucleotides had mean
lengths on the order of a few hundred nucleotides\footnote{Various groups have measured the
persistence length of single-stranded DNA in high salt concentrations
to be between 0.75 nm and 1.5 nm~\protect\cite{achter,busta,tinland}, or
roughly 1 to 2 nucleotides, meaning that the
polymers used were of order 100 persistence lengths long.}\ and were assumed to
be close to monodisperse.  After
adding the polynucleotides,
\bran\ monitored the transmembrane ionic current as a function of
time.  The time series shows a baseline current, modulated by periods on the
order of hundreds of microseconds in which the current decreases
almost to zero (figure~\ref{exptal-data}, inset).  A variety of observations support the interpretation
that these blockades were caused by the passage of a polymer through
the \alysin\ channel.  The data of \branc\ can thus be interpreted as
giving measurements of the times required for individual
polynucleotides to traverse the membrane under the influence of an
electric field.

\par
When these data are displayed as a histogram, with the number of
observed events plotted against the length of the blockade
(figure~\ref{exptal-data}), one sees that the blockade times fall into
three distinct peaks.  Of these, the first (``peak 1'') is caused by
polymers that enter and retract and thus do not completely cross the
membrane, while the other two (``peak 2'' and ``peak 3'') are both the
result of a polymer's actually passing through the channel.  The
polymers in peak 3 evidently cross the membrane roughly three times faster
than those in peak 2.
\bran\ made the intriguing suggestion that there are two
characteristic times associated with translocation because the
polynucleotide can enter the pore in two distinct directions: One peak
corresponds to polymers that enter the channel with their 3' end
first, the other to polymers that enter with their 5' end first.  We
will show in subsequent sections how such behavior can arise from a
simple microscopic model.

\par
A quantity of considerable interest in what follows will be the mean
force $F$ driving the polymer through the pore.\footnote{One can
define $F$ more precisely as the mean force required to immobilize a
given monomer in the pore, where the average is taken over time and
over all of the monomers in a given polymer.  Thus $F$ does not
include hydrodynamic drag forces, nor forces that vanish when averaged
over all the monomers.  Equivalently, $F$ can be defined by requiring
that $\exp(Fa/\kt)$ be the ratio of the probabilities that the polymer
will move forward one base to the probability that it will move
backwards one base, again appropriately averaged over all monomers.}
Clearly $F$ is primarily the result of the electric field acting on
the polymer.  Since a long, narrow channel has a much larger
electrical resistance than the macroscopic volumes of solution on
either side of the membrane, any voltage $V$ applied to the system
should fall almost entirely across the
\alysin\ pore.  The charge on each nucleotide is just the electron
charge $e$, so the electrostatic energy gained by moving one
nucleotide through the pore is $e V$.  This suggests that $F$
is roughly
\be
F \approx \frac{e V}{a} \approx 5 \frac{\kt}{a} \; , \label{f-defn}
\ee
where $a \approx 6 \AA$ is the length of a nucleotide, and the second
equality holds for $V \approx 125 {\rm mV}$.  For most of the rest of
the paper, we will assume that $F$ takes this value, and many of our
arguments will be based on the fact that $F$ is thus quite large when
expressed in appropriate units.  Some effects that could modify $F$
are considered in the discussion section and in
appendix~\ref{drive-app}.

\par
Before presenting our model, we would finally like to review
the experimental evidence that the interactions between the polymer
and the \alysin\ pore do indeed play the dominant role in KBBD's
experiments.  We have already mentioned the existence of two distinct
characteristic times for the polymer to cross the membrane.  Such a
result is easiest to interpret if one believes that the polymer's
speed is determined by interactions between the polymer and the narrow
channel constriction, where molecular scale asymmetries could be
important.  Similarly, recent data show that homo-polynucleotides of
different bases can move at strikingly different speeds (Dan Branton,
Harvard University,
personal communication): poly[U] is of order 20 times faster than
poly[dA].  Although chemical differences certainly can lead to
variations in polymer properties such as the persistence length, we
believe that such strong dependence on molecular details can more
easily be explained if we focus on the pore region.  Finally, even the
fastest polynucleotides pass through the pore far more slowly than
simple estimates of hydrodynamic drag would suggest: Model the pore as
a cylindrical hole of radius $R$ and the part of the polymer in the
pore as a cylinder of radius $r$.  Then, when the polymer moves with
speed $v$, the drag force per length on the part in the pore is
roughly $2 \pi \eta r v/(R-r)$.  Electrophoretic effects change this
result very little (see appendix~\ref{drive-app}).  For a
polynucleotide in an
\alysin\ channel, $r/(R-r)$ is somewhat larger than unity, and the
total length of the ``cylinder'' is roughly $50 {\rm \AA}$.  According
to scaling arguments of Lee and Obukhov~\cite{lee-obu}, the
contribution to the drag force from the ends of the polymer outside
the channel is only $2
\times 6 \pi
\eta b v$, where $\eta$ is the solvent viscosity, and the Kuhn length
$b$ is between 15 and $30 {\rm \AA}$.  Even if
hydrodynamic interactions are entirely screened by the motion of
counterions (as they are for the electrophoresis of an isolated
polymer in solution, with screening length of order the monomer size),
the drag on those parts of the polymer in solution cannot be larger
than roughly $4 \pi
\eta L v$.  If one substitutes typical parameter values for KBBD's
experiments and balances the sum of these drag forces with the naive
driving force of $5 \kb T$ per nucleotide, one finds that the polymer
would be expected to move through the pore at a rate of roughly $10^8
{\rm nucleotides}/{\rm second}$, 100 times faster than observed.  The
three observations of this paragraph, taken together, certainly
suggest that we focus on the degrees of freedom in the pore when
trying to understand the experiments of KBBD.

\section{Coarse-Grained Description}
\label{macro-model}
\subsection{Motivation and Governing Equation}

This section, and most of the rest of the paper, is concerned with
predicting distributions of blockage times of the sort shown in
figure~\ref{exptal-data}.  It is now well-established in condensed
matter physics that the form of the slow, long length-scale dynamics
of a system is often determined by the system's symmetries and
conservation laws.  All microscopic details are subsumed in
phenomenological coupling constants and transport coefficients.  In
this spirit, we would like to obtain a coarse-grained equation for the
probability $P(x,t)$ that a contour length $x$ of the polymer's
backbone has passed through the pore at time $t$. (The variable $x$ is
defined so that if the polymer backbone has length $L$, $x=0$ when the
polymer has just started in the pore and $x=L$ when it has reached the
other side).  For such a ``hydrodynamic'' description to make sense,
several conditions must be met.  One is that the polymer length $L$ be
much larger than the distance $a$ between successive nucleotides.  We
also demand that the dissolved counterions (as well as the solvent and
any other solutes ) relax quickly compared to the translocating
polymer, in order that we may ignore their dynamics.  Since the ions
are much smaller than a polynucleotide, and consequently diffuse much
faster, this condition should not be difficult to satisfy.  Finally,
our task will be considerably simplified if the microscopic system is
(approximately) invariant under translations by an integer multiple of
$a$ in either direction. Then, after averaging over variations on the
scale of a single nucleotide, we must obtain a translationally
invariant equation.  We will give this
assumption a firmer basis in the next section.  Roughly, however, there
should be translational invariance when we can neglect the parts of
the polymer outside of the channel, and this in turn should be possible when
the interactions between the polymer and the pore are strong enough.

\par
Under the
conditions just outlined, the (probability) density of the polymer is
the only conserved variable, and it is relatively straight-forward to
write down the coarse-grained hydrodynamic equation for $P$.  Because
there is only a single polymer (or, equivalently, a ``gas'' of
non-interacting polymers going through the same hole), the probability
current $j$, defined by $\partial P/\partial t + \partial j/\partial x =
0$ must be linear in $P$.  The lowest order allowed terms are then
proportional to $P$ and to $\partial P/\partial x$:
\be
j(x,t) = v P(x,t) - D \frac{\partial P(x,t)}{\partial x} \; .
\ee
The first term is permitted because there is an electric field driving
the system.  $P$ then satisfies the familiar equation for diffusion
with drift,
\be
\frac{\partial P}{\partial t} = D \frac{\partial^2 P}{\partial x^2} -
v \frac{\partial P}{\partial t} \; . \label{effect-diffn}
\ee
Here $v$ and $D$
are respectively an average drift velocity and an effective diffusion
coefficient.  Their values are determined by more microscopic physics;
in particular, they may depend nonlinearly on the applied electric
field.  Eq.~\ref{effect-diffn} may alternatively be derived from
a microscopic master equation that is invariant under translations by
$a$.  The coefficients $v$ and $D$ are then related to the lowest-lying
eigenvalues of the master equation.  This connection will be
illustrated in a subsequent section.

\par
On the macroscopic level of Eq.~\ref{effect-diffn},
 all information on the competition between driving and diffusive
spreading is encoded in a parameter that we call the diffusive
length $\ld \equiv D/v$.  Roughly speaking, on length scales less than
$l_d$, the polymer's motion is little affected by the presence of the
bias from the electric field, while on scales larger than $l_d$, the
driving dominates.  Indeed, if Eq.~\ref{effect-diffn}
described a rigid particle diffusing in one dimension under the
influence of a uniform force $f$, an Einstein relation would hold, and
we would have $v = D f/(\kb T)$, and $\ld = \kb T/f$.  Thus, in this case, $l_d$ is
precisely the length over which the driving force does a quantity
$k_{\rm B} T$ of work.  In the remainder of this section, we will
often assume that the length $L$ of the polymer is larger than $l_d$, a
condition satisfied by KBBD's data.

\subsection{Distribution of Passage Times}
\par
We now propose to calculate a distribution of passage times of the
sort measured by KBBD.  This section will show that, for given $v$ and
$D$, the probability $\psi(t)$ that the polynucleotide takes a time
$t$ to pass through the channel has only one peak.  Thus, the presence
of two peaks in KBBD's data must be explained by the assumption that
different physical situations give rise to different values of $v$ and
$D$.  Subsequent sections will argue that a polynucleotide passing
through the pore with its 3' end first can indeed have an average
velocity that is significantly different from one passing through with
its 5' end first.  This section, however, is confined to the
calculation of the passage times for fixed parameter values.  The
distribution $\psi(t)$ we obtain should thus be compared to a single
peak in the data of KBBD.

\par 
One can easily estimate the first few cumulants of this
distribution.  If a polymer of length $L$ moves
with average velocity $v$, one expects that the mean time to pass through
the channel should be $\tbar \approx L/v$.  Likewise,
the variance in the
distance traveled in a time \tbar\ is $(\Delta x)^2 = 2 D \tbar$.  It
would then seem reasonable that the variance in arrival times
should be $\bdelt^2 \equiv \langle (t-\tbar)^2 \rangle
\approx (\Delta x)^2/v^2$, or $\bdelt^2
\approx 2 D L/v^3$.  These conclusions are in
fact roughly correct for a sufficiently long
polymer.  One might expect corrections, however, because some fraction
of the polymers that enter the pore will leave again from the same
side instead of passing all the way through.  On average, these will
be the ``slower'' molecules: Those that spend a significant time with
only the tip of the polymer inserted in the channel are far more
likely to fall back out than are those that are quickly driven through the
hole.  Thus, only ``faster'' chains tend to enter into the calculation
of the mean transit time, decreasing
\tbar.  This effect is most pronounced for small $L/\ld$, because only
molecules within $l_d$ of the \cis\ side have an appreciable chance of
``backing out'' instead of exiting on the \trans\ side.  Indeed, when
$L \ll \ld$, the driving should be negligible, and we expect \tbar\ to
approach its $v=0$ value $L^2/6 D$.  To determine the precise form of
this crossover, we must turn to a more detailed calculation.

\par
This calculation can be formulated as one of a well-studied class of
problems known as first-passage problems~\cite{risken,vank}.
Essentially, all that is required is to solve Eq.~\ref{effect-diffn}
on the interval $[0,L]$ with absorbing boundary conditions $P(0) =
P(L) = 0$.  Then, the current density $j(L)$ at $L$ gives the
probability per time that the polymer will leave the pore from the far
({\em trans}) side, while $-j(0)$ is the probability per time that it
will exit from the {\em cis} side from which it entered. 
One must also specify the starting point
$x_0 \in [0,L]$ of the polymer; in what follows, we always take the
limit $x_0 \rightarrow 0$, in keeping with the fact that the polymer
starts entirely on the \cis\ side of the membrane.  The algebraic
details of the solution are summarized in appendix~\ref{fp-app}; here we
include only a discussion of the main results.

\par
Although exact expressions for \tbar\ and \bdelt\ may be obtained, it
turns out to be more instructive to consider the
distribution $\psi(t)$ itself.  For arbitrary $L/\ld$, this can only
be expressed as an infinite series, but if terms
that become exponentially small as $L^2/(v t \ld)
\rightarrow \infty$ are neglected, a comparatively simple analytic
expression is obtained:
\be
\psi(t) \simeq  \frac{v}{2} \sqrt{\frac{\ld}{\pi}} \left( \frac{L^2}{\ld
(v t)^{5/2}} - \frac{2}{(v t)^{3/2}} \right) e^{-(v t - L)^2/(4 v t
l_d) }. \label{psi-lim}
\ee
Note that this expression is not valid for sufficiently large $t$, and
in particular not for $t$ so large that it predicts that $\psi(t)$
becomes negative.  Nonetheless, for values of $t$ near the maximum in
$\psi(t)$, i.e. those such that $v t/L \sim {\cal O}(1)$, it is
accurate to within a percent for $L/\ld$ as small as 4, and correctly
reflects the qualitative features of $\psi(t)$ for significantly
smaller $L/\ld$.  Figure~\ref{fig1} plots $\psi(t)$ for $L/\ld = 5$; a
Gaussian with the same mean and variance is included for comparison.
Evidently, $\psi(t)$ is quite skewed, and its
mean and maximum are correspondingly well-separated.  Thus, \tbar\ and
\bdelt\ are not the best parameters
for describing experimental data.  Indeed, both cumulants are sensitive to how
$\psi(t)$ decays for large $t$, making them very hard to extract
accurately from realistic data sets.  A more useful choice of
parameters to characterize $\psi(t)$ are the position \tm\ of its
maximum (which satisfies $d
\psi/dt |_{\tm} = 0$) and the width \delt\ of the peak.  The latter
is defined as $\delt \equiv (t_R-t_L)/2$, where $t_R$ and $t_L$ satisfy
$\psi(t_R,t_L) = e^{-1/2}
\psi(\tm)$; we have chosen a factor of $e^{-1/2}$ instead of the more
conventional $1/2$ to facilitate comparison with fits of data to a
Gaussian.  One expects that as $L/\ld \rightarrow \infty$, \tm\ and
\delt\ should approach \tbar\ and \bdelt, respectively.  For example,
for large $L/\ld$ we have,
\be
\tm = \frac{L}{v} \left(1 - 5 \frac{\ld}{L} + \frac{17}{2}
\frac{\ld^2}{L^2} + 32 \frac{\ld^3}{L^3} + \cdots \right) \; .
\ee
The rapidly growing coefficients indicate that although \tm\ approaches
$L/v$ as $L$ approaches infinity, it falls away
from its asymptotic form quite rapidly for finite $L$. 

\par
More generally, one can easily find \tm\ and \delt\ by numerically
solving the equations that define them.  Figure~\ref{fig2} plots
$\delt/\tm$ versus the polymer length $L$.  This ratio is especially
interesting because it depends only on $L/\ld$, and not on $v$ and $D$
separately; one can thus use it quickly to estimate $L/\ld$.  In
KBBD's data,
$\delt/\tm$ is usually of order $0.5$ for a $\sim \! 200$ nucleotide
chain, suggesting that $L/\ld \approx 5$, or that $l_d$ is of order 40
nucleotides.  As figure~\ref{fig3} indicates, in this range \tm\ already
deviates significantly from the naive guess $\tm \approx L/v$.  In
particular, $\tm/L$ varies by a factor of 2 as $L/\ld$ increases from
5 to 25.  With sufficiently good data, this deviation from a strict
proportionality to $L$ might well be observable, providing strong
confirmation of our quasi one-dimensional picture.

\section{Regime of Validity}
\label{reg-valid}
\par
In the previous section, we argued that a requirement for the validity
of a one-dimensional diffusion model was that the
system be (approximately) unchanged if the polymer moves an integer
number of monomers forwards or backwards in the pore.  This section
discusses when this condition is satisfied.  We begin by dividing the
polymer into three parts: the roughly ten nucleotide long piece
that is actually {\em inside} the channel, and the two ``ends'',
comprising the majority of the nucleotides, {\em outside} the channel.
The pore always contains the same number of bases, so, for the
homopolymers, this part of the polymer always satisfies the
requirement of translational symmetry.  The length of each end
``dangling'' outside the pore, in contrast, changes with the
translocation parameter $x$, destroying translational invariance.  In
what follows, we shall argue that under certain conditions this
variation may be neglected.  Our arguments assume that the parts of
the polynucleotide outside the pore may be described by the theories
usually applied to long, flexible polymers~\cite{deg-bk,doi-ed}; we
thus ignore, for example, hydrogen-bonding and other specific
interactions~\cite{cant-sch}.  We also assume that the ion channel is
sufficiently long and narrow that any voltage drop falls almost
entirely across the channel (see appendix~\ref{drive-app}).  The electric field
and the solvent flow velocity outside of the channel can then be
ignored.

\par
There are two criteria for ignoring the ends of the polymer outside of
the pore.  First, they should have a characteristic relaxation time
that is much faster than the characteristic time for the motion of a
monomer through the channel.  In the absence of interactions between
the polymer and the pore, one would expect diffusion on the scale of a
few monomers to be much faster than the relaxation of a long polymer
coil, and this inequality could never be satisfied.  However, since
the nucleotides in the pore can be expected to interact strongly with
the confining protein, the requirement is not implausible.  The
longest time scale of an isolated polymer in solution is the Zimm time
$\tz \approx 0.4 \eta R_G^3/(\kb T) \approx 0.4 \eta N^{3
\nu} b^3/(\kb T)$, where $\nu$ is the Flory exponent,\footnote{In
principle, $\nu \approx 0.6$ for a long polymer in a good solvent.
However, even with the longest available chains, $\nu$ is never
observed experimentally to be larger than 0.55 \protect\cite{doi-ed},
so we use this value for specific numerical calculations.}  $b$ is the
Kuhn segment length (equal to twice the persistence length), $\eta$ is
the solvent viscosity, and $N = L/b$.  Substituting in numerical
values for a single-stranded polynucleotide in water, one finds that
$\tz
\approx N^{3 \nu} (3.2
\times 10^{-4} \, \mu{\rm sec})$.  If we imagine that the polymer moves a monomer
through the channel by hopping over an energetic barrier (an idea to
be considered in more detail when we introduce our microscopic model),
then in the limit of strong driving, the translocation speed is simply
$v = a/\tp$, where \tp\ is the longest relaxation time of the part of
the polymer in the pore.  Substituting numerical values for poly[U],
we find $\tp = a/v \approx 1.5 \mu{\rm sec}$.  Comparing this figure
to \tz, we see that the two become of the same order when $N$ is of
order 150, corresponding to a length of polymer of roughly 300
nucleotides protruding from each side of the pore.  Of course, for
polymers that traverse the membrane more slowly, as is the case for
poly[dA], the value of $N$ above $\tz \gta \tp$ can be significantly
larger.

\par
As long as the dynamics of the polymer outside of the pore are fast
compared to the dynamics in the pore, one need not treat the external
degrees of freedom explicitly.  Instead, they affect the motion of the
polymer only through a contribution $\cf(x)$ to its free energy and
through the increased drag they contribute.\footnote{Here, we assume
that $v$ is sufficiently small that the parts of the polymer outside
the pore are essentially in equilibrium.  On purely dimensional
grounds, this must be true when $\tz \ll N^y (b/v)$ for some
non-negative exponent $y$, a requirement that is met in KBBD's
experiments.}  Lee and Obukhov's scaling argument~\cite{lee-obu}
implies that their effect on the drag is independent of the length of
polymer on a given side of the membrane.  On the other hand, in order
for us to be able to neglect
\cf, $d\cf/dx$ must be small compared to the force $F$ driving
translocation.  Denote the free energy of the coil on the
\cis\ side of the membrane by $\cf_{\rm C}(x)$ and that of the coil on
the \trans\ side by $\cf_{\rm T}(x)$; their sum is $\cf(x)$.  Sung and Park
pointed out that $\cf_{\rm C}$ and $\cf_{\rm T}$ are simply the free
energies of a polymer grafted by one end to a planar
surface~\cite{sung-park}.  For a polymer of length $x$, this entropic
free energy is known
to be proportional to $\kb T \ln(x/b)$, with a coefficient of order
unity that depends on whether excluded volume effects are
important~\cite{binder}.  Ignoring the few monomers
actually in the channel, the lengths of polymer on the \cis\ and
\trans\ sides of the barrier are $x$ and $L-x$, respectively, so
\be
\cf(x) \propto \kb T \, \left[ \ln\left(\frac{x}{b}\right) + \ln\left(\frac{L-x}{b}\right) \right] \; .
\label{cal-f}
\ee
For a chain that is a fixed fraction of the way through the hole
(i.e. for fixed $x/L$), $d\cf/dx$ vanishes like $1/L$.  Further, it
makes little sense to consider $x<a$, where $a$ is the length of a
single monomer, so we must always have $d\cf/dx
\lta \kb T/a$.  Typical values will be much smaller than this bound.
The driving force $F \approx 5 \kt/a$ thus greatly exceeds $d\cf/dx$;
indeed, since the polymers used by KBBD are several hundred
nucleotides long, $F$ is more than a factor of 100 larger than a
typical value of $d\cf/dx$.
In sum, we have shown that in the window of polymer lengths
\be
\frac{\kt}{F a} \ll N \ll \left(\frac{\kt a}{\eta b^3
v}\right)^{1/3\nu}
\ee
 the polymer is
short enough to relax quickly, but long enough that the entropic
barrier to crossing the membrane is not too steep.  For lengths in
this window, the ends of the chain hanging outside of the pore can be
neglected compared to the monomers inside the pore.  Since the system
studied by KBBD falls within this window, we are justified in using
simple one-dimensional models to describe it.

\section{Microscopic Model of the Pore}
\label{micro-model}
\par
Until now, we have avoided specifying the physics of the interactions
within the pore.  In this section, we present a simple
phenomenological model of these interactions.  Our main goal is to
understand physically how the parameters $v$ and $D$ can vary
sufficiently to explain experimental facts like the difference
in velocities between polymers moving forwards and backwards.

\subsection{Description of the Model}
\par
We begin by focusing on the polymer backbone, whose coordinate $x$
tells us what fraction of the polymer chain has passed through the
channel.  If the motion of the backbone is sufficiently slow compared
to all the other degrees of freedom in the pore, then we can take $x$
to be the only dynamical variable in the problem.  The remaining
degrees of freedom are then described by a free energy $\Phi(x)$ that
depends on the polymer translocation parameter $x$.  The potential
$\Phi(x)$ can, for example, be expected to have contributions from
electrostatic interactions between the polymer and the \alysin\
heptamer.  Two unit charges separated by $1 {\rm \AA}$ in water have
an energy of about $6 \kb T$ at room temperature; since both
polynucleotide and protein have completely ionized groups in
physiological pH, it is thus plausible that typical values of $\Phi$
should be at least on the order of several $\kb T$. We split $\Phi$
into a mean slope $F$ determined by the applied voltage drop and a
part $U(x)$ that captures the details of the polymer's interactions
with the pore: $\Phi(x) = U(x) - F x$.\footnote{In principle, $U$
could depend on the applied voltage (and hence on $F$).  We ignore
this effect; many of our conclusions will in any case turn out to be
insensitive to it.}  For homopolymers (provided we continue to neglect
the degrees of freedom outside the pore), $U(x)$ is periodic, with
period $a = 1 \, {\rm nucleotide}$.  $F$ is precisely the mean force
introduced in Eq.~\ref{f-defn}, and is equal to $eV/a$ in the
simplest picture.

\par
Our problem is now formally no
different from that of a point particle diffusing in a periodic
potential $U$ and driven by a constant force $F$.  The probability $P(x)$ of
finding such a particle at a point $x$ is governed by a
Smoluchowski equation,
\bea
\frac{\partial P}{\partial t} & = & D_0 \frac{\partial}{\partial x} \left[
\frac{\partial P}{\partial x} + \frac{U'(x) - F}{\kt} P \right]  \nonumber \\
& \equiv & \lfp P \; .
\label{main-diffn}
\eea
The ``bare'' diffusion constant $D_0$ is related through an Einstein
relation to some suitable drag force on the polymer in the channel.
It is not to be confused with the ``effective'' diffusion constant $D$
that includes the effects of $U$ and describes the polymer's motion on
length scales much larger than $a$.

\par
It is helpful both for numerical work and for intuition
building to have a concrete idea of the simplest form $U(x)$ could
take.  In particular, such a simplified ``cartoon'' will give us an
idea of the minimum number of parameters needed to describe the gross
features of the potential.  A natural choice for such a $U(x)$ is a
sawtooth potential of the sort sketched in figure~\ref{fig4}.  It is
described by two dimensionless parameters, the peak height $U_0/\kt$
and the asymmetry parameter $\alpha$.  When $\alpha = 1/2$, the
potential is perfectly symmetrical, while $\alpha = 0$ or 1
corresponds to maximal asymmetry.  In addition to $U(x)$, the full
potential $\Phi$ contains a term proportional to the driving force
$F$, which figures in the dimensionless group $F a/\kt$.  Thus, to
specify our potential fully, we require the three dimensionless
parameters $U_0/\kt$, $\alpha$, and $F a/\kt$, as well as $D_0$ and
the repeat distance $a$, which set a time and a length scale.  More
generally, we expect that any form of $U(x)$ with only one peak
per period will be roughly characterized by a peak height $U_0$ (equal
to the difference between the minimum and the maximum values of
$U(x)$), and an asymmetry $\alpha$ (defined as the distance between a
minimum in $U(x)$ and the next maximum to the right, divided by $a$).
While we have no {\em a priori} information about $\alpha$, we have
suggested that $U_0$ should be of order several $\kb T$, and have
argued $Fa/\kt
\approx 5$ for KBBD's experiments.

\par
Because the time required to
diffuse over a barrier depends exponentially on the barrier height,
small differences in $U_0$ can lead to significant changes in
translocation speed, consistent with KBBD's observations.  
Further, if $U(x)$ is asymmetrical,
forces $F$ and $-F$ will lead to different barrier heights, and thus
to different mean drift speeds for the diffusing polymer.
Figure~\ref{fig5} illustrates this point.  It thus appears plausible
that a polymer moving through the pore with its 5' end first could
travel at a very different speed from one going through the pore 3'
end first.  There is, however, one additional complication.  As
shown in figure~\ref{fig6}, three different ``vector'' quantities can
be oriented relative to the membrane: the applied electric field, the \alysin\
pore, and the DNA.  Each can point towards the \cis\ or the \trans\
chamber.  With, say, the pore orientation held fixed, there are thus
four possible situations.
The two that have been realized in the
experiments of KBBD are related by a flip of the polymer, while
transforming $F \mapsto -F$ (or equivalently $U(x) \mapsto U(-x)\;$)
in our model amounts to changing the direction of the field.  Thus,
although a velocity difference of a factor of 3 can clearly be
accounted for by reasonable choices of parameters, we cannot
quantitatively address how the two peaks in KBBD's experiments are
related.  Once all four possible situations have been explored experimentally,
however, it should be possible, for example, to estimate the value of
$\alpha$ by comparing data for the appropriate pairs of situations.

\subsection{Effective Mobility and Diffusion Coefficient}
\label{mobility-calcn}
\par
We now turn to the task of calculating the parameters $v$ and $D$ that
describe the behavior of Eq.~\ref{main-diffn} on long length scales.
Several approaches are available; in this section, we will describe
the results of an analysis based on ideas of Risken~\cite{risken}.
Details of the calculation, which relies on an eigenfunction expansion,
are given in appendix~\ref{vd-app}.  In the most general case, $v$ and
$D$ have fairly complicated forms, but relatively simple limiting
cases capture most of the relevant behavior.  For example, one
finds~\cite{led-vinokur,scheidl}
\be
\frac{1}{v} = \frac{1}{D_0} \int_0^{\infty} dz e^{-Fz/\kt} \int_0^a
\frac{dx}{a} e^{[U(x+z)-U(x)]/\kt} \;  \label{exact-veloc},
\ee
from which a number of limiting behaviors can be extracted.  Several
equivalent expressions for $v$, as well as a similar, but more
involved, expression for $D$, can also be obtained.

\par
Figure~\ref{fig7} plots the velocity $v$ versus $F$ for polymers
traveling in two different directions in the same (asymmetric)
potential.  At typical values of $F$, differences in velocity between
forwards and backwards motion of a factor of 3 or more are easily
obtained.  Likewise, the calculated velocities are much slower than
they would have been in the absence of a potential.

\par
One can gain more quantitative insight into both of these observations
by studying how $v$ and $D$ behave in various limiting cases.
Relegating the derivations to appendix~\ref{vd-app}, we next consider
several such expressions.  Three cases are particularly of interest:
large and small driving force $F$, and large potential barriers $U_0$
(the case of small $U_0$ corresponds to the absence of a potential and
was discussed earlier).  For small $F$, $v$ and $D$ must satisfy
an Einstein relation.  Indeed, in this limit one finds,
\bea
v & = & \frac{D_0 F}{\kt}\frac{1}{\ione \itwo}\left[1 + {\cal O}(\frac{F a}{\kt}) 
\right] \nonumber \\
D &=& D_0 \frac{1}{\ione \itwo}\left[1+ {\cal O}(\frac{F a}{\kt})\right]
  \; , \label{asympt1}
\eea
where 
\bea
\ione & =& \int_0^a\frac{dx}{a} e^{U(x)/\kt} \nonumber \\
\itwo & = &
\int_0^a \frac{dx}{a} e^{-U(x)/\kt} \; .
\label{ionetwo}
\eea
  Thus, $v/D = F/\kt$, as the fluctuation-dissipation theorem
requires, but the effective diffusion coefficient $D$ is reduced from
its bare value $D_0$ by a factor that grows exponentially with the
characteristic height of the potential.  Perhaps more surprising is
the fact that a linear-response-like regime is also reached for
sufficiently large $F$.  As $F \rightarrow \infty$,
\bea
v & = & D_0 F \left[1+{\cal O}(\frac{U_0}{F a})^2\right] \nonumber \\
D &=& D_0
\left[1+{\cal O}(\frac{U_0}{F a})^2\right] \; . \label{asympt2}
\eea
The physical content of this result is that when $F$ is much larger
than a typical force derived from $U(x)$, $\Phi'(x) \approx -F$, and
contributions from $U$ may be neglected entirely.  In the opposite
limit of large $U_0$, one might expect that the diffusion process can
essentially be described as hopping from one potential minimum to the
next.  Approximate formulas based on the Kramers escape
rate~\cite{vank} should then apply.  In fact, for large $U_0$ one
finds
\be
v \simeq \frac{D_0}{a \ione \itwo} \left[e^{\alpha F a/\kt} - e^{-(1-\alpha)Fa/\kt}
\right]
\ee
and
\be
 D \simeq \frac{D_0}{2 \ione \itwo} \left[e^{\alpha
F a/\kt} + e^{-(1-\alpha) F a/\kt} \right] \; . \label{asympt3}
\ee
As before,
we select the origin of $U(x)$ so that its maximum and minimum in each
period occur at points $x_{\rm max} > x_{\rm min}$, with $ x_{\rm max}
- x_{\rm min} = \alpha a$. 

\par
We have already estimated from KBBD's data that $\ld \equiv D/v
\approx 40 a$.  A striking feature of the asymptotic forms
Eq.~\ref{asympt1} through Eq.~\ref{asympt3} just obtained is that all
three imply a much smaller value.  As we noted when we introduced the
parameter $l_d$, the linear response results both yield $\ld = \kt/F$;
given our naive estimate $Fa/\kt \approx 5$, we find $\ld
\approx a/5 \ll 40 a$.  For $U_0$ large enough that the ``hopping''
approximation of Eq.~\ref{asympt3} applies, this order of magnitude is
little changed even as $F \rightarrow \infty$.  Indeed, in this limit
Eq.~\ref{asympt3} gives $\ld = a/2$.  It is of course possible that
some particular form of $U(x)$ with finite $U_0$ and $F$ might lead to
a value of $l_d$ of order $40 a$.  It seems more likely, however, that
$l_d$ interpolates reasonably smoothly among its various limiting
values.  The inset to figure~\ref{fig7} illustrates this point for the
sawtooth potential introduced earlier.  Although $v$ and $D$ each
separately can depend strongly on the shape of $U(x)$, their ratio is
far less sensitive.  We are thus led to one of the central conclusions
of this paper: While many aspects of KBBD's results can be
qualitatively explained by a model of diffusion in a one-dimensional
periodic potential, the observed width of their peaks is inconsistent
with this model if one takes $F a \approx 5 \kt$.

\section{Discussion}
\label{discussion}

\par
In the previous section, we argued that the peaks in KBBD's
distribution of first passage times are much wider than is consistent
with our minimal one-dimensional model.  It is not difficult to
suggest reasons why this might be the case.  Perhaps the most obvious
is that $F a/\kt$ could differ significantly from 5.  Not only would a
decrease of a factor of 100 in $F$ bring our prediction for $l_d$ into
line with experimental observations, it would also explain the
polymer's unexpectedly slow translocation speed.  At least two effects
might decrease $F$.  First, unless the pore has infinite resistance,
not all of the applied voltage drop $V$ will be across the pore.
Although the large ($\sim {\rm M}\Omega$) resistance of the
\alysin\ channel makes it unlikely that this mechanism could
diminish $F$ by orders of magnitude, it certainly leads to some
decrease.  Second, the fact that there is a nonzero ionic current
flowing through the pore while the polymer is translocating means that
the motion of the polymer itself need not satisfy detailed balance.
That is, the {\em error rate}, or ratio of the probabilities of moving
forward one base to moving backwards one base, is no longer required
to be equal to $\exp(eV/\kt)$.  To use a somewhat different language,
as the counterions are forced through the pore by the electric field,
they entrain some of the solvent along with them.  This solvent flow
exerts an additional drag force on the polymer, and this drag
contributes to the mean force $F$.  As a result, the electrophoretic
mobility of the polymer in the channel is not in general equal to its
hydrodynamic mobility multiplied by its charge.
Appendix~\ref{drive-app} presents simple estimates based on continuum
mechanics that suggest that both of these effects are small.  These
estimates, however, make a number of simplifications; indeed, even the
validity of the continuum equations is not assured on the nanometer
scale.  Given the importance of a large value of $Fa/\kt$ to any
attempts to sequence polynucleotides using the \alysin\ pore, it thus
seems desirable to verify experimentally that it is indeed
approximately 5.

\par
Although a smaller than expected driving force is certainly one
mechanism that would generate wider peaks, others exist that do not
require a large error rate.  In many ways, our most poorly justified
assumption is that the motion of the polymer backbone through the pore
is much slower than the relaxation of every other degree of freedom in
the system, so we begin by considering what might happen if this
assumption were to break down.  For example, the protonation state of
the open \alysin\ channel is known to fluctuate on a much slower time
scale than the characteristic polymer time $a/v
\sim 1 \mu{\rm sec}$~\cite{kasia}, and the energy barrier to moving a
base through the pore might change significantly when the protonation
state changes.  It is instructive to consider a naive extension of
our one-dimensional model meant crudely to describe such a situation.
Suppose that the pore + polymer system can be in one of two states,
state 1, in which the polymer backbone can diffuse freely, and state
2, in which the backbone is trapped and cannot move.  Let there be
a transition rate (per time) $\omega_{ij}$ from state $i$ to state
$j$.  This situation bears some similarities to popular models of
motor proteins~\cite{julicher}, but with the important difference that
the ratio $\omega_{12}/\omega_{21}$ need not violate detailed balance;
a similar description has also recently been proposed for the
one-dimensional motion of RNA polymerase along a
polynucleotide~\cite{jul-bruinsma}.
If $P_i(x)$ is the probability that the system is in state $i$ and
that a length $x$ of polymer has passed through the pore, the long
time diffusion of the system is governed by equations of the form
\bea
\frac{\partial P_1}{\partial t} &  = & D_1 \frac{\partial^2 P_1}{\partial
x^2} - v_1 \frac{\partial P_1}{\partial x} - \omega_{12} P_1 +
\omega_{21} P_2 \\
\frac{\partial P_2}{\partial t} & = & \omega_{12} P_1 - \omega_{21} P_2 \; .
\eea
In state 2, the motion is arrested, so both the velocity and the
diffusion coefficient vanish.  Just as in the one-dimensional,
periodic case (appendix~\ref{vd-app}), this model leads to a spreading
Gaussian wave packet, with velocity and diffusion coefficient
determined by the behavior of the eigenvalues near zero.  One finds a
velocity
\be
v = \frac{v_1}{2} \left(1+\frac{\Delta \omega}{\omega} \right) \; ,
\ee
and a diffusion coefficient
\be
D = \frac{D_1}{2} \left(1+\frac{\Delta \omega}{\omega} \right) +
\frac{v_1^2}{2 \omega} \left(1 - \frac{(\Delta \omega)^2}{\omega^2} \right) \; ,
\ee
where $\Delta \omega = \omega_{21}-\omega_{12}$ and $\omega =
\omega_{12}+\omega_{21}$.  Thus, if
$v_1 \neq 0$ and $\omega$ and $\Delta \omega$ are chosen properly,
$\ld = D/v$ can be made arbitrarily large.  This is true even if
$D_1/v_1$ remains of order $a$.Thus, even in this simple example,
broad peaks are possible as soon as one relaxes the constraint that
the model only contain one degree of freedom.

\par
 In addition to the possibility that there is more than one slow
degree of freedom, many other factors could contribute to the wide
blockage time distributions observed by KBBD.  For example, the
observed peaks could reflect not the distribution in passage times for
polymers of a given length, but the length distribution of the
polymers themselves.  Although the polydispersity of the poly[U]
samples used by KBBD was not well characterized, no qualitative
differences were seen with a perfectly monodisperse sample of DNA (Dan
Branton, Harvard University, personal communication), suggesting that
polydispersity is not the culprit.  Similarly, we calculated the free
energy ${\cal F}$ of the parts of the polynucleotide outside the pore
using a model that applies to conventional polymers above the theta
point.  If ${\cal F}$ took a different form, we might not be able to
neglect it.  In particular, significant asymmetries between the two
sides of the membrane could result in a non-electrical contribution
to the driving force~\cite{park-sungb,dimarzio,carl}.  For example, if
the polynucleotide adsorbs weakly on one side of the membrane, there
would be a force towards the adsorbing side of order $f \kt/a$, where
$f$ is the fraction of adsorbed monomers.  A similar effect could be
obtained by confining the polymer on only one side of the membrane.
Indeed, for sequencing applications, it might be useful intentionally
to introduce an asymmetry as a way to manipulate the polymer's speed
and error rate without affecting the ionic current.  Similarly, the
current can in principle be varied with little effect on the polymer
by putting a high concentration of macroions (e.g. colloidal particles)
on one side of the membrane.  This would induce a concentration
gradient in their counterions that tended to drive the ions across the
membrane.

\par
Finally, we would like to touch on one other issue of particular
relevance to efforts to sequence polynucleotides as they pass through
the pore.  All of our results up to this point have been strictly
valid only for homopolymers.  Since it is known that diffusion in
random media can be qualitatively different from diffusion in ordered
systems~\cite{bouch-georges}, it is worth asking whether we expect any
important changes when the homopolymer is replaced by a random
heteropolymer.  As long as the assumptions leading to our model of
diffusion in a one-dimensional, periodic medium hold, one can argue
that the effect of using heteropolymeric DNA would be to modify the
potential $U(x)$.  Rather than having an identical form within each
unit cell of length $a$, $U(x)$ might take one of four different
shapes, corresponding to four different bases.  It is known that
one-dimensional diffusion becomes anomalous when
$\overline{[U(x)-U(0)]^2} \rightarrow \infty$ as $x
\rightarrow \infty$, where the overbar indicates an average
with respect to the random distribution of
bases~\cite{bouch-georges,led-vinokur,scheidl}.  In biological DNA
sequences, it is believed that successive bases are either uncorrelated or have
correlations that decay algebraically with distance~\cite{herzel,stanley}.  In
either case, $\overline{[U(x)-U(0)]^2}$ remains bounded for large $x$.
Thus, within the simple one-dimensional model, no qualitatively new
behavior would result from replacing homopolymers with
heteropolymers.  This is what one would expect to observe for short
polynucleotides like those used by KBBD.  It is worth mentioning,
however, that this conclusion is sensitive to other effects.  For
example, the electrophoretic mobility of the polymer in the pore presumably has
some sequence dependence; this would lead to an effective short-range
correlated random {\em force} on the polymer.  Such a random force,
however small, would in principle result in anomalous diffusion on
sufficiently long length scales~\cite{bouch-georges,fisher-led}.

\section{Conclusion}

\par
The central idea of this paper was first presented in the
introduction: In the experiments of KBBD, and likely in other examples
of the translocation of biopolymers, the channel through which the
polymer passes cannot be viewed simply as a set of hard, homogeneous
walls.  Rather, more specific interactions between the polymer and the
channel must be taken into account.  Indeed, we have argued that there
is a regime in which polymer-pore interactions dominate, allowing a
quasi-one-dimensional description of the translocation process.  One
immediate consequence of this observation is that on long enough
length scales, the transport of the polymer through the pore is
governed by a simple phenomenological equation.  Starting from this
equation, we have derived several predictions about the polymer's
distribution of passage times.  For example, we have shown that the
polymer's mean translocation time depends linearly on its length only
for an extremely long polymer.  It is perhaps worth reemphasizing that
none of these results depend on any particular microscopic model of
the pore.  In contrast, several important qualitative observations of
KBBD can be understood in terms of a more microscopic picture
involving a ``tilted washboard'' potential.  The tilted washboard
model also lead us to point out that the distribution of passage times
in KBBD's experiments was far broader than one might expect from
simple estimates, or indeed from any model with only one degree of
freedom.  We have suggested several ways that this discrepancy might
arise.  Some, such as a serious mis-estimate of the mean force $F$ on
the polymer, imply an error rate in DNA sequencing of almost 50\%.
Several others, however, do not require revision of the estimated
error rate.  Most notable among these are polydispersity in the
polymer lengths and a strong coupling between the polymer and another
degree of freedom in the pore with slow dynamics.  Determining which
mechanism is at work in the experiments of KBBD remains an important
experimental and theoretical challenge.

\par
Our conclusions suggest several experimental avenues that might be
explored in KBBD's or some analogous system.  Most obvious would be to
try to measure the error rate (or equivalently, the driving force
$F$).  At least a rough estimate of $Fa/\kt$ might be obtained in
several ways.  Obviously, any experiment in which it is possible to
detect the passage of a particular nucleotide through the pore gives
one direct access to the error rate.  Alternatively, at small enough
applied voltage $V$, it should be possible to observe a linear
response regime.  Deviations from linear behavior would then be
observed at a value of $V$ such that $F a/\kt
\sim {\cal O}(1)$.  With linear response data from all four possible
relative orientations, it should also be possible to observe that
$dv/dV$ is the same for pairs of peaks.  More ambitiously, if one
could exert a non-electrical force on the polymer strong enough
that its mean velocity through the pore fell to zero, this would give
a direct measurement of $F$.  Such an experiment might be
accomplished with modern micromanipulation techniques.

\par
Other experiments of interest might test the existence of a
quasi-one-dimensional regime.  With enough data on the length
dependence of \tm, for example, it should be possible to observe the
predicted deviation from the simple guess $\tm \propto L$.  Further,
if this data could be extended to sufficiently long polymers,
deviations from the curve of figure~\ref{fig3} would provide
information on the crossover to a regime in which the dynamics of the
polymer outside the pore are slower than those inside the pore.  Once
the basic theory has been verified, a number of different directions
remain open.  For example, a study of the fluctuations of the ionic
current with a polymer in the pore could provide evidence about
whether there are important slow degrees of freedom other than the
polymer backbone itself.  Adding a time varying component to the
applied voltage in the experiments of KBBD might provide advantages in
sequencing applications; the characteristic frequency for motion from
one base to the next appears to be a relatively low $10^6$ Hertz.  The
behavior of polymers in very narrow channels is a rich subject that
has only begun to be investigated.

\setcounter{secnumdepth}{1}   

\appendix

\section{Details of Calculation of Distribution of Blockage Times}
\label{fp-app}
\par
We wish to solve the equation
\be
\frac{\partial P}{\partial t'} = \ld \frac{\partial^2
P}{\partial x^2} - \frac{\partial P}{\partial x} \;,
\ee
where $t' = vt$, for $P(x,t')$ subject to the boundary conditions that
$P$ vanish at $x=0,L$ and the initial condition $P(x,t'=0) =
\delta(x-x_0)$.  The right and left eigenfunctions of $\ld
\partial^2/\partial x^2 - \partial/\partial x$ are respectively
$\exp(x/2\ld) \sin(k_n x)$ and $\exp(-x/2\ld) \sin(k_n x)$, so we have
\be
P(x,t') = \frac{2}{L} \sum_{n = 1}^{\infty} e^{-\lambda_n t'}
 e^{-x_0/2 \ld} \sin(k_n x_0) e^{x/2\ld} \sin(k_n x) \; ,
\ee
with $k_n = n \pi/L$ and $\lambda_n = \ld k_n^2 + 1/4\ld$.  After
taking a derivative with respect to $x$, setting $x=L$, and
simplifying, one finds that the probability that the polymer will exit
the channel at $x=L$ at time $t'$ is
\bea
\varphi(t') = j(L)& = &  -\frac{2 \ld}{L} e^{(L-x_0)/2 \ld} \nonumber \\
&& \times \sum_n e^{-\lambda_n t'} k_n \sin[k_n (x_0-L)] \; .
\eea
Note that for large $L$ this is a very slowly convergent series.  It
is thus convenient to rewrite it using the Poisson resummation
formula.  This formula states that for any function $f$,
$\sum_{n=-\infty}^{\infty} f(n) = \sum_{m=-\infty}^{\infty} \hat{f}(2
\pi m)$, where $\hat{f}$ is the Fourier transform of $f$ defined by
$\hat{f}(q) = \int dx f(x) e^{i q x}$.  Rewriting the sum in terms of
the Fourier transform of the summand moves $L$ from the denominator to
the numerator of the exponential, yielding, after a little algebra,
\bea
\varphi(t') & = & \frac{1}{2 \sqrt{\pi \ld t'^3} } e^{(L-x_0)/2\ld}
e^{-t'/4\ld} \nonumber \\
& & \times \sum_{n=1,3,5,\ldots} e^{-(n-1)L/2\ld} \nonumber \\
& &\times \left[ (n L-x_0) e^{-(t'-nL+x_0)^2/4\ld t'} \right. \nonumber \\
& &\left. \mbox{}- (n L+x_0) e^{-x_0/\ld}
e^{-(t'-nL-x_0)^2/4 \ld t'} \right] \; .
\eea
With $\varphi$ written in this form, the first term is exponentially
larger than all subsequent terms as $L$ becomes large, facilitating
the analysis of limiting cases.  With the help of a few definite
integrals a number of results can be obtained exactly.  In
particular, if one defines $I(\alpha) = \int_0^\infty dy/y^{\alpha}
\exp[-(y-1)^2/4 \gamma y]$, then $I(3/2) = I(1/2) = 2 \sqrt{\pi
\gamma}$.  The first equality can be proven with the substitution $z =
1/y$, the second with the substitution $u = (y-1)/\sqrt{y}$;
$I(\alpha)$ for other values of $\alpha$ may be obtained by succesive
integrations by parts.  Using these identities, one can show, for
example, that the total probability that the polymer will exit from the
\trans\ side ($x=L$) if it started at $x_0$ is
\bea
\int_0^{\infty} dt' \varphi(t')
&  =& \left(1-e^{-x_0/\ld}\right)
\sum_{n=1,3,5,\ldots} e^{-(n-1)L/2\ld} \nonumber \\
& = &
\frac{1-e^{-x_0/\ld}}{1-e^{-L/\ld}} \; . \label{prob-trans}
\eea
One can similarly obtain exact expressions for $\langle t' \rangle$
and for higher moments.

\par
Thus far, we have allowed the polymer's starting point $x_0$
to be arbitrary.  Since the polymer always starts entirely on the
\cis\ side of the channel, the case of interest to us is $x_0
\rightarrow 0$.  Eq.~(\ref{prob-trans}) shows that the
probability that the polymer passes through the pore vanishes in this
limit.  This conclusion is, however, a pathology of our model.  We can
still obtain a meaningful {\em conditional} distribution of passage times
(that is, a distribution of passage times for those polymers that do
leave at $x=L$) by normalizing $\varphi$ by the total probability of
passage.  In the limit $x_0 \rightarrow 0$, one obtains
\bea
\psi(t') & = & \lim_{x_0 \rightarrow 0} \left[
\frac{1-e^{-L/\ld}}{1-e^{-x_0/\ld}} \varphi(t') \right] \nonumber \\
& = & 2 \sqrt{\frac{\ld}{\pi t'^3}} \left(1-e^{-\frac{L}{\ld}}\right)
\nonumber \\
& & \times \!\! \sum_{n=1,3,5,\ldots} \left(\frac{n^2 L^2}{\ld t'} -2 \right)
e^{-\frac{(n-1)L}{2 \ld}} e^{-\frac{(t'-nL)^2}{4 \ld t'}} .
\eea
All the terms but the first are subdominant as $L\rightarrow \infty$,
and when they are dropped, we obtain~(\ref{psi-lim}).  Note that not
only is Eq.~\ref{psi-lim} the correct asymptotic form for large
$L$, but also that all subsequent terms describe peaks centered at
increasingly larger values of $t'$.  Thus, even when these terms
significantly modify the behavior of $\psi(t')$ as $t' \rightarrow
\infty$, they can have a very small effect in the vicinity of \tm.

\section{Details of Calculation of Mobility and Diffusion Coefficient}
\label{vd-app}
\subsection{Exact Expressions}
\par
In this appendix, we will derive the expression of
Eq.~\ref{exact-veloc} for the mean drift velocity $v$ of a particle in
a periodic potential and an analogous expression for the effective
diffusion coefficient $D$.  Since the linear operator \lfp\ (defined
by Eq.~\ref{main-diffn}) is periodic, it must have eigenfunctions of
the Block form $\psinr = e^{i k x} \unr$, where $|k| < \pi/a$ and
\unr\ is periodic with period $a$.  The eigenfunction \psinr\ is
defined by $\lfp \psinr = -\lam \psinr$.  Since \lfp\ is a
non-hermitian operator, right and left eigenfunctions are not equal,
so we distinguish between them with superscripts R and L. Likewise,
the eigenvalues \lamtext\ are not in general real.  The eigenfunctions are
labelled by a band index $n$ and a wavevector $k$ in the first
Brillouin zone.  If the polymer
starts at $x= x_0$ at $t=0$, then $P(x,t)$ may be expressed as an
eigenfunction expansion:
\be
P(x,t) = \sum_n \int_{\rm BZ} dk \, e^{i k (x-x_0)} \unlo^* \unr
e^{-\lam t} \; . \label{eigen_P}
\ee
Because of the exponential decay $e^{-\lam t}$, the smallest values of
${\rm Re}\, \lam$ determine the behavior of $P$ at long times.  One can
prove that the lowest value occurs at $\lambda_0(k=0) = 0$.
Performing a saddle point integration about this point, one thus finds
that as $t \rightarrow \infty$,\footnote{Note that this result is
valid only for values of $x$ such that the difference $x-x_0-v t \ll
{\cal O}(\sqrt{t})$ for large $t$, so that the quantity in the
exponential has a saddle point exactly at $k=0$ instead of somewhere
in the upper complex $k$ half-plane.}
\be
P(x,t) \simeq \frac{1}{\sqrt{4 \pi D t} } e^{ -(x-x_0-v t)^2/(4 D t)}
\statold \; ,
\ee
where
\be
v = -i \left. \frac{d \lambda_0}{d k} \right|_{k=0} \; \; \; {\rm and} \; \;
\; D = \left. 2 \frac{d^2 \lambda_0}{dk^2} \right|_{k=0} \; .
\ee
At long times $P(x,t)$ is thus a spreading Gaussian, modulated by a
periodic function \statold\ that gives detailed structure on the scale
of $a$.  One can hence reasonably interpret the constants $v$ and $D$
in the expression for the Gaussian envelope as the same constants that
appear in the macroscopic Eq.~\ref{effect-diffn}.

\par
In light of these expressions, we obviously want to study the behavior
of $\lambda_0$ in the vicinity of $k=0$.  To do this, it is convenient
to rephrase the eigenvalue condition $\lfp \psinr = \lam \psinr$ in
terms of \unr\ as
\bea
 &&D_0 \left(\frac{\partial}{\partial x} + i k\right)
\left[\left(\frac{\partial}{\partial x} + i k\right) + \frac{\Phi'(x)}{\kt}
\right] \unr \nonumber \\
&& \; \; = D_0 \left[ \lfp + i k \left(2 \frac{\partial}{\partial x} +
\frac{\Phi'(x)}{\kt} \right) -k^2 \right] \unr \nonumber \\
&&\; \; =  -\lam \unr \; .
\eea
If we view the $k$-dependent part of the operator on the left as a
small perturbation on \lfp, then finding the derivatives of
$\lambda_0(k)$ at $k=0$ is formally the same as a problem in
quantum-mechanical perturbation theory.\footnote{There is a slight
difference in that in our case the perturbation to the original
operator \lfp\ has terms of order $k^2$ as well as of order $k$.}  As
usual, we pose the expansions
\bea
\lambda_0(k) & = & k\left.  \frac{d \lambda_0}{dk} \right|_{k=0}+
\frac{k^2}{2}\left. \frac{d^2 \lambda_0}{dk^2} \right|_{k=0} +
\cdots \nonumber \\
u_0^{\rm R}(k,x) & = & \ur_0(0,x) + k u_0^{\rm R,1}(x) + \cdots \nonumber \\
\ul_0(k,x) & = & \ul_0(0,x) + k u_0^{\rm L,1}(x) + \cdots \;
. \label{expan}
\eea
The ``ground state'' $\ur_0(k=0,x)$ of
the unperturbed problem can be obtained exactly by integrating $\lfp
\ur_0(k=0,x) = 0$~\cite{risken}:
\be
\statold = N e^{-\Phi(x)/\kt} - \frac{S}{D_0} \int_0^x dx' \,
e^{\Phi(x')/\kt} \; .
\ee
The two constants of integration $N$ and $S$ are determined by
normalization and by requiring that \statold\ be periodic.
Physically, $S$ is the (constant) probability current density in the
stationary state \statold.  Since $\lfp^{\dag} = D_0 \left[
\frac{\partial}{\partial x} - \frac{\Phi'}{\kt} \right]
\frac{\partial}{\partial x}$, $\ul_0(k=0,x)$ is clearly a constant,
which we choose to be 1.  The first correction to $\lambda_0(k=0)$ is
then just the ``expectation value'' of the perturbation in the
``ground state'':
\be
\left. \frac{d \lambda}{d k} \right|_{k=0} = i v = i \int_0^a dx
\left( 2 \frac{\partial}{\partial x} + \frac{\Phi'(x)}{\kt}
\right) \statold \; .
\ee
After a bit of algebra, one obtains
\be
v = S a = D_0 a \frac{\mydel}{a^2 [I_1 I_2 - \mydel I_3]} \;,
\label{veloc-risken}
\ee
where
\bea
I_1 & = & \int_0^a \frac{dx}{a} \, e^{\Phi(x)/\kt} \; , \\
I_2 & = & \int_0^a \frac{dx}{a} \, e^{-\Phi(x)/\kt} \; , \; {\rm and} \\
I_3 & = & \int_0^a \frac{dx}{a} \int_0^x \frac{dx'}{a} e^{-\Phi(x)/\kt} e^{\Phi(x')/\kt} \; .
\eea
Note that the integrals \ione\ and \itwo\ defined in Eq.~\ref{ionetwo}
are just $I_1$ and $I_2$ evaluated with $F=0$.  Finally, $v$ may be
put in the form of Eq.~\ref{exact-veloc} by everywhere rescaling $a$
as $a \mapsto n a$ and letting $n \rightarrow
\infty$~\cite{led-vinokur,scheidl}.  This substitution is valid
because we could have originally thought of $U(x)$ as having period $n
a$ instead of $a$ for any positive integer $n$.

\par
To find the ${\cal O}(k^2)$ correction to $\lambda_0(k=0)$ requires
knowing the ${\cal O}(k)$ correction to the ground state \uoo.  This
is typically expressed as a sum of eigenfunctions of the unperturbed
problem, which are assumed known.  Instead, we will invert
\lfp\ by direct integration to find an analytic expression for \uoo.
After substituting the expansions of Eq.~\ref{expan} into the
eigenvalue equation and equating terms of order $k$, we find
\bea
\lfp \uoo & = & -iv \statold \nonumber \\
&& - i D_0 \left(2
\frac{\partial}{\partial x} + \frac{\Phi'(x)}{\kt}\right) \statold \; .
\eea
Everything on the right hand side of the equation is known, so \uoo\
can be found by integrating twice.  The constants of integration are
determined by demanding that \uoo\ be periodic and that its inner
product with $\ul_0(k=0,x) = 1$ vanish.  Once \uoo\ is known, we can
equate terms of order $k^2$ to learn that
\be
D-D_0 = \int_0^a \ul_0(k=0,x)^* \, i\left(2 \frac{\partial}{\partial x} +
\frac{\Phi'(x)}{\kt} \right) \uoo \; .
\ee
A certain amount of additional algebra finally leads to an expression
for $D$:
\be
\frac{D}{D_0} = 1 + L \frac{I_2 J_1 -\mydel J_2}{I_1 I_2 - \mydel I_3}
- J_3 \; ,
\ee
where the $I_i$ are the same as before,
\bea
J_1 & = & \int_0^a\frac{dx}{a} \int_0^x \frac{dx'}{a} \, e^{\Phi(x)/\kt} f(x') , \nonumber \\
&& \\
J_2 & = & \int_0^a \frac{dx}{a} \int_0^x \frac{dx'}{a} \int_0^{x'} \frac{dx''}{a} \, e^{-\Phi(x)/\kt}
e^{\Phi(x')/\kt} f(x'') \; , \nonumber \\
&& \\
J_3 & = & \int_0^a \frac{dx}{a} \int_0^x \frac{dx'}{a} \, f(x') \; ,
\nonumber \\
&&
\eea
and
\be
f(x) = \frac{S a}{D_0}\statold + \frac{\partial \statold}{\partial x} -
\frac{S}{D_0} \; .
\ee
Just as with the expression of Eq.~\ref{veloc-risken} for the velocity, we
can find equivalent formulas for $D$ by replacing $a$ by $n a$
and letting $n \rightarrow \infty$.

\subsection{Approximate Expressions}
\par
Having obtained exact expressions for $v$ and $D$, we will now outline
how the limiting forms of Eq.~\ref{asympt1} through Eq.~\ref{asympt3}
are derived.  We will focus on the expressions for $v$; the
manipulations required to obtain the analogous asymptotic forms for
$D$ are very similar.
\par
The behavior of $v$ for large and small $F$ is most easily
studied starting from Eq.~\ref{exact-veloc}.  Define $A(z) \equiv
\int_0^a dx/a
\exp[U(x+z)/\kt-U(x)/\kt]$, and note that $A$ is a periodic function of $z$
with period $a$.  Expanding $A(z)$ in a Fourier series and Laplace
transforming term by term as demanded by Eq.~\ref{exact-veloc} leads
immediately to the formula of Eq.~\ref{asympt1} for the small $F$
behavior; in particular, the leading behavior is determined by the
constant term in the series, $\int_0^a dz/a A(z) = \ione \itwo$.
Similary, we may find the behavior of $v$ for large $F$ by successive
integrations by parts: $D_0/v = A(0)/F + A'(0)/F^2 + \cdots$.
\par
The starting point for finding the large $U$ behavior is similar.
Rewrite $v$ as
\bea
\frac{D_0 a}{v}& = & \int_0^a dx \, e^{-U(x)/\kt} e^{F x/\kt} \nonumber \\
&& \times \int_x^{\infty} dy \,
e^{U(y)/\kt} e^{-F y/\kt} \; .
\eea
First the inner, and then the outer integral can then be evaluated by
Laplace's method as $U$ becomes large.  Note that since $U(x)$ is
periodic, we must sum a geometric series over an infinite number of
extrema to find the asymptotic expression for the inner integral.
Also, the location of the first maximum depends on the lower bound of
integration $x$.  Once the two integrals have been evaluated, an
expression equivalent to Eq.~\ref{asympt3} follows immediately.
The only difference is that in Eq.~\ref{asympt3}, we have chosen
to write \ione\ and \itwo\ instead of their large $U$ forms.

\section{Estimate of Reduction in the Driving Force}
\label{drive-app}
\par
In the discussion section, we mentionned two factors that should
reduce the driving force $F$ on the polymer from the naive value $Fa/\kt
\approx 5$.  In this appendix, we present order-of-magnitude estimates
of how large a correction these effects cause.  The estimates are
based on the equations of continuum mechanics in a simplified
geometry and use bulk parameter values.  They thus neglect a number
of subtleties, most notably the presence of significant numbers of
charged groups on the \alysin\ pore itself.

\par
We begin by looking at the additional drag on the polyelectrolyte due
to the flow of oppositely charged small ions through the pore.
Consider a cylindrical polymer of radius $r$ inside a cylindrical pore
of radius $R$, and assume that the distance $\delta = R-r$ between the
polymer and the pore satisfies $\delta \lta r$ and $\delta \lta
\kappa^{-1}$, where $\kappa^{-1}$ is the Debye-H\"{u}ckel screening
length.  Both conditions hold in KBBD's experiments.  Then neither the
steady-state density $n$ of ions nor the solvent velocity $u$ parallel
to the cylinder axis vary too strongly with radial distance.  In
particular, there is no Manning condensation.  In the presence of an
applied electric field ${\bf E}$ (of order the applied voltage $V$
divided by the length of the channel), a body force term $e n {\bf E}$
must be added to the Stokes equation that describes very viscous flow.
Here $e$ is the electron charge, and we have assumed that the only
ions present are monovalent cations; the figure we obtain for the
additional drag will thus be an upper bound on the drag possible with
ions of both signs.  We may use the component of the Stokes equation
along the cylinder axis to estimate
\be
e n E \sim \eta \frac{\ui}{\delta^2} \; ,
\label{approx-stokes}
\ee
where $\eta$ is the solvent viscosity.  Since the Stokes equation is
linear, we have written the axial solvent velocity as $u = \up + \ui$, where
\ui\ is driven by the electrical force on the ions, and \up\ by the
motion of the polymer.  The electrical current density $J$ due to the
ions is then roughly
\be
J \sim e^2 n \mu E + e n (\up + \ui) \; .
\label{approx-current}
\ee
Here $\mu$ is the mobility of a single counterion. The first term
describes motion of the counterions relative to the solvent, and the
second the convection of the counterions by solvent motion.  The value
of $J$ is set by the fact that the total current $I \sim 2 \pi r
\delta J$ is known to be of order 10 pA in KBBD's experiments; \up\
must be of order the polymer's translocation speed.  Thus, the problem
is reduced to solving the two ``equations''~\ref{approx-stokes}
and~\ref{approx-current} in the two unknowns $n$ and \up.  By
substituting reasonable parameter values, one finds that the term
proportional to $\up + \ui$ in Eq.~\ref{approx-current} may be
dropped.  Then, the extra drag force per length on the polymer is
\be
\frac{\rm drag}{\rm length} \sim 2 \pi r \frac{\eta \ui}{\delta} \sim
\frac{I}{\mu e}
\ee
Substituting a typical value of an ionic mobility (in bulk solution),
and multiplying by the channel's length $\lch \sim 50 {\rm \AA}$,
we obtain a drag force of order $10^{-7} {\rm dyne}$ due to the
ion-driven solvent flow.  This is to be compared to $5 \kt/a \approx 3
\times 10^{-6} {\rm dyne}$.  We conclude that the presence of the
counterions in the channel does not significantly reduce $F$.

\par
Another factor that we should take into account concerns the
assumption that any applied voltage drop $V$ falls entirely across the
channel.  In reality, some electric field must ``leak'' out of the
pore, decreasing the force that drives the polymer.  The simplest
model in which one can consider this problem assumes that the bulk
solution is neutral and has a constant conductivity $\sigma$, so that
the electrical current density and the electric field are related by
${\bf J} = \sigma {\bf E}$.  This picture of course does not hold
within a distance $\kappa^{-1}$ of the membrane, but should be
reasonably accurate on longer length scales. Focus only on one side of
the membrane, so that the pore acts as a current source injecting a
current $I$ at the boundary of an infinite half-space.  Far way from
the pore, the current density, and thus the electric field, should
decay like $1/r^2$.  In the near field, one expects this decay to be
cut off at a distance of order the pore radius $R$.  Knowing the
current $I$, we can immediately obtain the electric potential, and
thus the voltage drop $V_c$ across the channel, in terms of the voltage
$V$ imposed at infinity.  Our rough estimate gives (in cgs units)
\be
V - V_c \sim 2 \int_\infty^R \frac{I}{2 \pi \sigma r^2} dr \sim
\frac{I}{\pi \sigma R} \; .
\ee
The factor of 2 in front of the integral arises because we must
include the effect of a voltage decrease on both sides of the
membrane.  Substituting experimental values for $I$, $\sigma$, and
$R$, we find that this calculation suggests that $V_c$ differs from
$V$ by only a few percent.

\vspace{0.2in}

It is a pleasure to thank Dan Branton and Jene Golovchenko for
introducing us to this problem and for numerous helpful discussions.
We would also like to thank Armand Ajdari, Meredith Betterton,
Jean-Fran\c{c}ois Joanny, Didier Long, and Sergei Obukhov for
insightful comments and for pointing out useful references.  This work
was supported by the Harvard Materials Research Science and
Engineering Laboratory through Grant No. DMR94--00396 and by the
National Science Foundation through Grant No. DMR97--14725.  One of us
(D.K.L.) also acknowledges support from a National Science Foundation
Graduate Research Fellowship.

\newpage

\bfig
\caption{Histogram of number of observed blockade events versus the
lifetime of the blockade, for 210 nucleotide poly[U].  The numbers 1
through 3 label the different peaks.  From KBBD (courtesy of Dan
Branton, Harvard University).  {\em Inset:}\ Typical time series of
the current versus time in the experiments of KBBD, showing a
transient blockade due to the translocation of a polymer (courtesy of
Dan Branton, Harvard University).
\label{exptal-data}}
\efig

\bfig
\caption{The distribution $\psi(t)$ of passage times plotted
versus $t$ for $L/\ld = 5$.  Both quantities are appropriately
non-dimensionalized, $t$ as $v t/L$ and $\psi(t)$ as $L \psi(t)/v$.  The
dashed curve is a Gaussian with the same mean and variance as
$\psi(t)$. \label{fig1}}
\efig

\bfig
\caption{Plot of the relative width $\delt/\tm$ of the peak in the
distribution of passage times, versus $\ld/L$.  This curve may be used
to obtain the quick estimate $\ld \approx 40$ nucleotides for
the system studied by KBBD.  The dashed curve gives the $L \rightarrow
\infty$ asymptotic behavior, $\delt/\tm \sim \protect\sqrt{2 \ld/L}$.
We have chosen to put $\ld/L$ instead of $L/\ld$ along the ordinate to
allow smooth contact with this large $L$ behavior. \label{fig2}}
\efig

\bfig
\caption{ $v \tm /L$ plotted versus $L/\ld$.  Note that $v \tm/L$ varies
signficantly over the range of $L/\ld$ relevant to the experiments of
\bran, and in particular that it does not reach its asymptotic value
of unity until well outside the range of this plot.  {\em Inset\/}:
Plot of \tm\ (nondimensionalized by $\ld/v$) versus $L$
(nondimensionalized by $l_d$).  The dashed line gives the large $L$
limiting form $L/v$, the solid line the exact value.  Note
that although \tm\ appears to the eye to depend linearly on $L$ over
much of the range of the plot, it still differs significantly from
$L/v$.
\label{fig3}}
\efig

\bfig
\caption{Sketch of the sawtooth ``cartoon'' potential discussed in the
text.  The potential has period $a$, and $\alpha a$ is the distance
from one minimum to the next maximum.  The parameter $U_0$ gives the
energy difference between minimum and maximum.
\label{fig4}}
\efig

\bfig
\caption{Sketch showing how asymmetry in the potential can lead to
different speeds for forwards and backwards motion.  A bias is applied
to the unperturbed potential (A) at top so that it has the same average
gradient in the two bottom pictures.  The potential at right (B), however,
has been reflected through the vertical axis before the gradient is
applied.  It thus has larger barriers to hopping from one minimum to the
next than the potential at left (C), leading to slower dynamics.
\label{fig5} }
\efig

\bfig
\caption{The four possible relative orientations of polymer, pore, and
applied electric field.  In KBBD's experiments, the relative
orientation of the pore and field is fixed and the orientation of the
polymer is allowed to vary, corresponding to cases B and D.  In our microscopic
model, the pairs (A, B) and (C, D) are related by $F \mapsto -F$.
\label{fig6} }
\efig

\bfig
\caption{Plot of the (nondimensionalized) average velocity $v$ from
equation~\protect\ref{exact-veloc} versus the driving force $F a/\kt$;
$v$ is calculated using our microscopic model with a sawtooth
potential.  The parameter values are $U_0/\kt = 10$; $\alpha = 0.7$
for the upper curve and $\alpha = 0.3$ for the lower curve. The
potentials for the two curves are thus related by $U(x) \mapsto
U(-x)$. {\em Inset\/}: The diffusive length $l_d$ versus the
barrier height $U_0$ of the sawtooth potential, for fixed driving
force $F a = 5 \kt$ and asymmetry $\alpha = 0.7$.  Note that over the
entire range of $U_0$, $\ld \lta a$.
\label{fig7} }
\efig

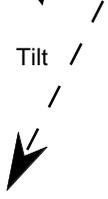
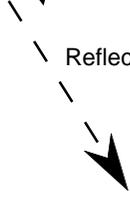
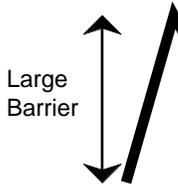
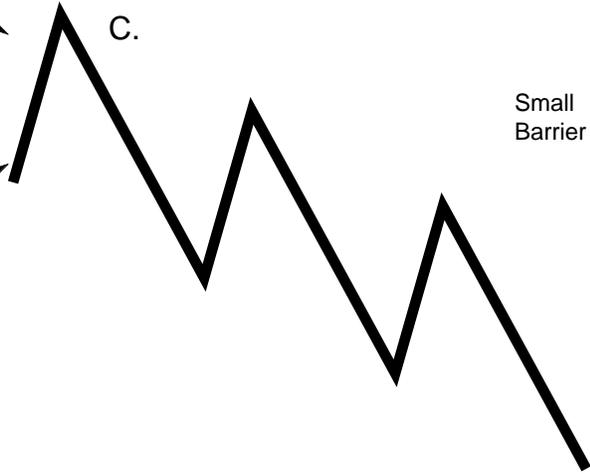
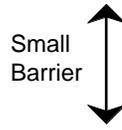
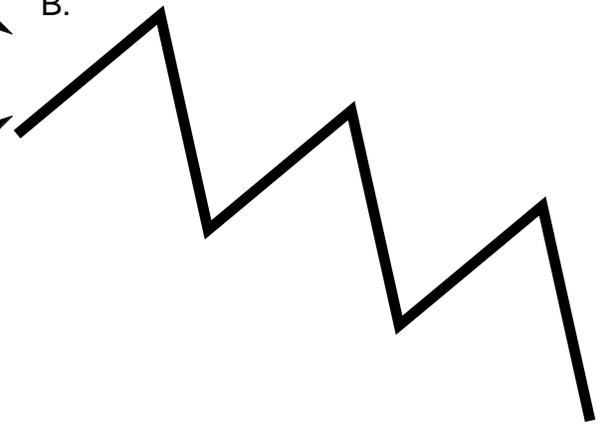
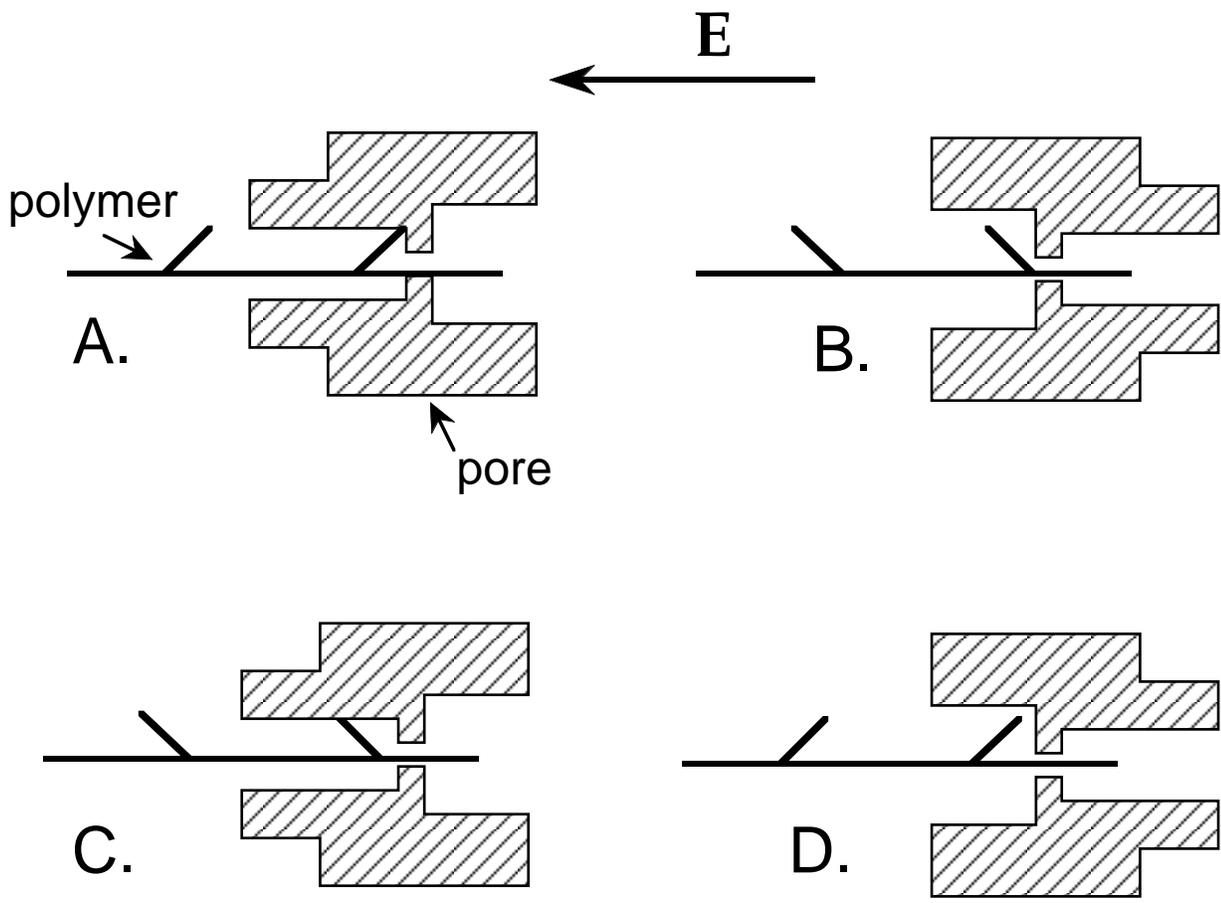
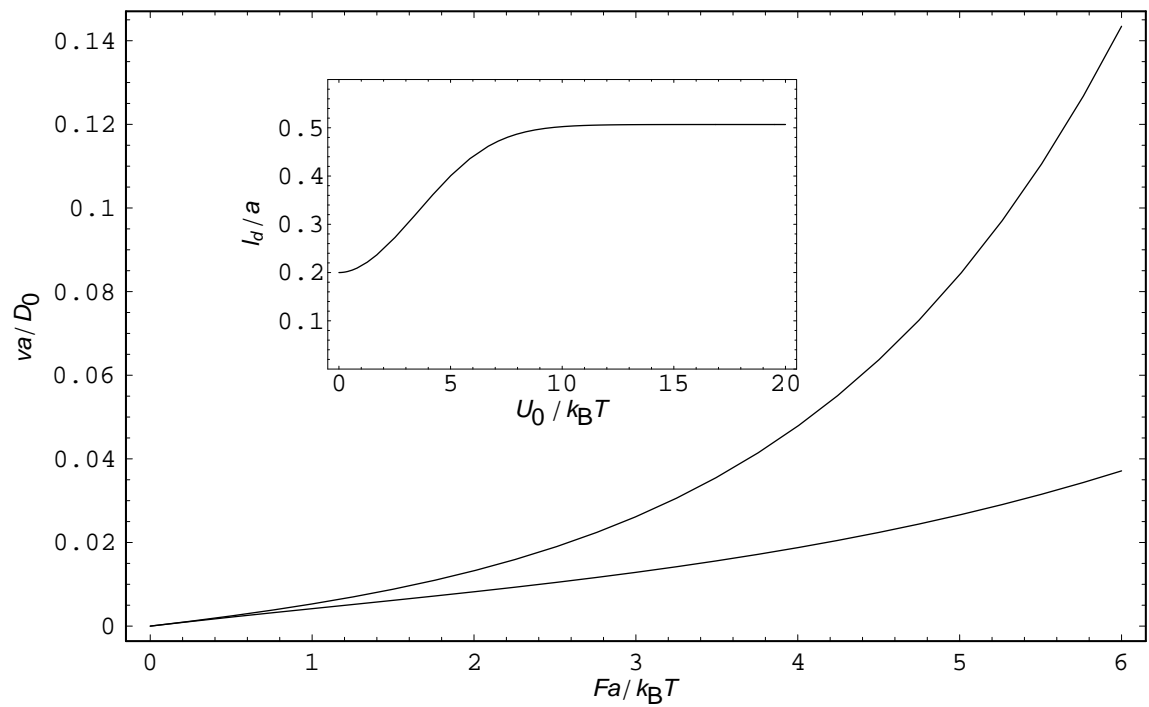
\end{document}